\RequirePackage{fix-cm}
\documentclass[smallextended]{svjour3}       
\smartqed  
\usepackage{graphicx}
\usepackage{amssymb}
\usepackage{amsmath}

\begin{document}

\title{Approximate Pricing of Derivatives Under Fractional Stochastic Volatility Model
}


\author{Yuecai Han         \and
        Xudong Zheng 
}


\institute{Y. Han \at
              School of Mathematics, Jilin University, Changchun, 130012, China \\
              \email{hanyc@jlu.edu.cn}           
           \and
           X. Zheng \at
              School of Mathematics, Jilin University, Changchun, 130012, China \\
                            \email{zhengxd20@ mails.jlu.edu.cn} 
}

\date{Received: date / Accepted: date}

\maketitle

\begin{abstract}
We investigate the problem of pricing derivatives under a fractional stochastic volatility model. We obtain an approximate expression of the derivative price where the stochastic volatility can be composed of deterministic functions of time and fractional Ornstein-Uhlenbeck process. Numerical simulations are given to illustrate the feasibility and operability of the approximation, and also demonstrate the effect of long-range on derivative prices.
\keywords{ Pricing derivatives \and Stochastic volatility \and Fractional Brownian motion}
\subclass{60G22 \and 91G20}
\end{abstract}

\section{Introduction}
\label{intro}

The Black-Scholes-Merton pricing formula is the cornerstone of derivatives analysis\cite{1,2}. However, the stochastic volatility models describe the actual patterns of financial markets better than the model with constant volatility, because the data in financial markets show that the volatilities fluctuate continuously. Johnson and Shanno use the Monte Carlo method to price a European call option with the stochastical variance\cite{4}. Wiggins numerically solves the call option valuation problem under a general continuous stochastic process for return volatility\cite{6}. Chernov et al. evaluate the role of various volatility specifications, such as multiple stochastic volatility factors and jump components, in the appropriate modeling of equity return distributions\cite{3}. Stein and Stein study the stock price distribution that arises when prices follow a diffusion process with a stochastical volatility parameter(i.e. Stein-Stein model)\cite{7}.

For continuous sampling, Neuberger uses a nonparametric approach to study Delta hedging strategies based on variance swaps under log contract\cite{9}. Dupire investigates a stochastic volatility model whereby traders in exotic options can extend the Black-Scholes model of option pricing\cite{10}. These methods hold for arbitrary stochastic volatility processes, so the replication does not require the assumption of a particular stochastic volatility model. In other words, the impact of stochastic volatility is already reflected through the prices of the options used in the replication.

In terms of discrete sampling, Elliott et al. solve the pricing problem of swaps by probabilistic methods and partial differential equation methods\cite{11}. Sepp analyzes the effect of discrete sampling on the valuation of options on the realized variance in the Heston stochastic volatility model. As a result, short-term options on the realized variance are priced by the semi-analytical Fourier transform methods\cite{12}. Bernard and Cui investigate the fair strike of a discrete variance swap for the general time-homogeneous stochastic volatility model, and their result yields asymptotics for the discrete variance swaps\cite{13}.

With regard to numerical simulation, Little and Pant investigate the pricing problem of swaps by using the finite element approach\cite{15}. Windcliff et al. discuss the valuation and hedging of discretely observed volatility derivatives and use a numerical approach to solve the partial differential equations satisfied by the forward prices of volatility derivatives\cite{16}.

All the above stochastic volatility models are driven by Brownian motion. However, Mandelbrot and Ness find a long-range correlation in the returns of stocks in financial markets\cite{17}. Beben and Orłowski illustrate the long-time correlations in both well-developed and emerging market indexes\cite{19}. Huang and Yang illustrate that there is some long-term memory for various data frequencies and lags in the UK market\cite{21}. In other words, the future price of the underlying asset is not only related to the present price but also the price over a considerable period in the past. 

Mandelbrot and Ness propose fractional Brownian motion as a process based on the path integral form of standard Brownian motion\cite{17}. Subsequently, Lin investigates the $L^{2}$ integration theory on the fractional Brownian motion\cite{24}, but Rogers then proves that there are arbitrage opportunities in the financial model, which makes it unreasonable to price derivatives in this type of financial market\cite{25}. Duncan et al. define the multiple and iterated integrals of a fractional Brownian motion and give various properties of these integrals\cite{26}. Elliott and Hoek present an extended framework for fractional Brownian motion in which processes with all indices can be considered under the same probability measure\cite{29}. Biagini et al. introduce the theory of stochastic integration for fractional Brownian motion based on white-noise theory and differentiation\cite{28}. These works present a framework for fractional Brownian motion in which processes with all indices can be considered under the same probability measure. As an application, Necula generalized the risk-neutral valuation pricing formula in the framework of fractional Wick-type integrals\cite{27}. Contrary to the situation when the pathwise integration is used, Hu and Øksendal proved that the fractional Black–Scholes market has no arbitrage if using the stochastic integration developed by Duncan et al.\cite{23}. 

Gatheral et al. propose that log volatility behaves essentially as a fractional Brownian motion with Hurst parameter $H\in\left( 0,1\right) $, at any reasonable timescale. The concept of rough fractional stochastic volatility (RFSV) is also given and they demonstrate that the RFSV is remarkably consistent with financial time series data\cite{33}. Bayer et al. show how the RFSV model can be used to price claims on both the underlying and integrated variance. In particular, they find that the rBergomi model fits the SPX volatility markedly better than conventional Markovian stochastic volatility models, and with fewer parameters\cite{32}. Cheridito et al. propose the fractional Ornstein-Uhlenbeck process by proving that the Langevin equation with fractional white noise also has a stationary solution\cite{31}. Garnier and Sølna present an analysis for the case when the stationary stochastic volatility model is constructed in terms of a fractional Ornstein-Uhlenbeck process\cite{34}. However, some classical models (e.g. the Stein-Stein model and the Heston model) introducing fractional Brownian motion have more general requirements for the volatility form. We propose an approximate pricing method for the fractional stochastic volatility model by solving stochastic partial differential equations with variable coefficients, where the volatility is constructed by a deterministic function of time and the fractional Ornstein-Uhlenbeck process. 

The paper is organized as follows. In section \ref{sec:basics}, we introduce some basic background on the fractional Brownian motion and the fractional Ornstein-Uhlenbeck process. Our main results are in section \ref{sec:main}. We derive the approximate pricing formula and prove that the approximation error can be limited. In section \ref{example}, we calculate the price of the European option under the fractional Stein-Stein model as an example to illustrate the feasibility and operability of the method. Numerical simulations are presented in section \ref{simulations} as well as the conclusions in section \ref{sec:conclusions}.

\section{Fractional Brownian motion and Ornstein-Uhlenbeck process}
\label{sec:basics}

\subsection{Fractional Brownian motion}
\label{sec:basics_1}
The fractional Brownian motion with Hurst parameter $H\in\left( 0,1\right) $ is a zero-mean Gaussian process $\left ( B_{t}^{H} \right )_{t \in \mathbb{R} } $ with covariance
\begin{displaymath}
	\mathbb{E}\left[B_{t}^{H} B_{s}^{H}\right]=\frac{\sigma_{H}^{2}}{2}\left(|t|^{2 H}+|s|^{2 H}-|t-s|^{2 H}\right),
\end{displaymath}
where
\begin{displaymath}
	\begin{aligned}
		\sigma_{H}^{2} &=\frac{1}{\Gamma\left(H+1/2\right)^{2}}\left[\int_{0}^{\infty}\left((1+s)^{H-1/2}-s^{H-1/2}\right)^{2} d s+\frac{1}{2 H}\right] \\
		&=\frac{1}{\Gamma(2 H+1) \sin (\pi H)}.
	\end{aligned}
\end{displaymath}

We use the following moving-average stochastic integral representation of the  fractional Brownian motion\cite{22}
\begin{displaymath}
	B_{t}^{H}=\frac{1}{\Gamma\left(H+1/2\right)} \int_{\mathbb{R}}(t-s)_{+}^{H-1/2}-(-s)_{+}^{H-1/2} d B_{s},
\end{displaymath}
where $\left ( B_{t} \right )_{t \in \mathbb{R} } $ is a standard Brownian motion. Moreover, the filtration $\mathcal{F}_{t}$ generated by $B_{t}^{H}$ is also the one generated by $B_{t} .$ 

The fractional Brownian motion is self-similar, i.e.$ \left ( B_{at}^{H},t \in \mathbb{R} \right)  $ and $  ( a^{H} B_{t}^{H},$ $t \in \mathbb{R}  ) $ have the same probability law for all $a>0$. Compared to Brownian motion, it displays a long-range dependence and positive correlation properties when $1/2<H<1$  and it displays negative  correlation property when $0<H<1/2$. This special property of fractional Brownian motion allows it to describe models with path-dependent. To prove that the results of this paper apply to the fractional Stein-Stein model, the following Itô formula under fractional Brownian motion is needed when the Hurst parameter is strictly larger than $1/2$.

Let $\phi: \mathbb{R}_{+} \times \mathbb{R}_{+} \rightarrow \mathbb{R}_{+}$be given by
$$
\phi(s, t)=H(2 H-1)|s-t|^{2 H-2} .
$$
For $p\ge 1$, let $L^p(\Omega, \mathcal{F}, P)=L^p$ be the space of all random variables $F: \Omega \rightarrow \mathbb{R}$ such that
$$
\|F\|_p:=\left(\mathbb{E}|F|^p\right)^{1 / p}<\infty
$$
and let $L_\phi^2\left(\mathbb{R}_{+}\right)=\left\{\left.f\left|f: \mathbb{R}_{+} \rightarrow \mathbb{R},\right| f\right|_\phi ^2:=\int_0^{\infty} \int_0^{\infty} f_s f_t \phi(s, t) d s d t<\infty\right\}$. For notational simplicity, $L_\phi^2\left(\mathbb{R}_{+}\right)$ is denoted by $L_\phi^2$.

Let $\Phi$ be given by
$$
(\Phi g)(t)=\int_0^{\infty} \phi(t, u) g_u d u,
$$
where $g \in L_\phi^2$. The $\phi$-derivative of a random variable $F \in L^p$ in the direction of $\Phi g$ where $g \in L_\phi^2$ is defined as
$$
D_{\Phi g} F(\omega)=\lim _{\delta \rightarrow 0} \frac{1}{\delta}\left\{F\left(\omega+\delta \int_0(\Phi g)(u) d u\right)-F(\omega)\right\}
$$
provided the limit exists in $L^p(\Omega, \mathcal{F}, P)$. Furthermore, if there is a process $\left(D^\phi F_s, s \geq 0\right)$ such that
$$
D_{\Phi g} F=\int_0^{\infty} D^\phi F_s g_s d s \quad  a.s. 
$$
for all $g \in L_\phi^2$, then $F$ is said to be $\phi$-differentiable.

$D_s^\phi$ is defined as a Malliavin directional derivative. Let $\mathcal{L}(0, T)$ be the family of stochastic processes on $[0, T]$ such that $F \in \mathcal{L}(0, T)$ if $\mathbb{E}|F|_\phi^2<\infty, F$ is $\phi$-differentiable, the trace of $(D_s^\phi F_t$, $0 \leq s \leq T$, $0 \leq t \leq T)$ exists, $\mathbb{E} \int_0^T\left(D_s^\phi F_s\right)^2 d s<\infty$, and for each sequence of partitions $\left(\pi_n, n \in \mathbb{N}\right)$ such that $\left|\pi_n\right| \rightarrow 0$ as $n \rightarrow \infty$, the quantities
$$
\sum_{i=0}^{n-1} \mathbb{E}\left\{\int_{t_i^{(n)}}^{t_{i+1}^{(n)}}\left|D_s^\phi F_{t_i^{(n)}}^\pi-D_s^\phi F_s\right| d s\right\}^2
$$
and $\mathbb{E}\left|F^\pi-F\right|_\phi^2$ tend to 0 as $n \rightarrow \infty$, where $\pi_n: 0=t_0^{(n)}<t_1^{(n)}<\cdots<t_{n-1}^{(n)}<t_n^{(n)}=T$.

\begin{lemma}[\cite{26}, Theorem 4.3]\label{thm:1}
	Let $\eta_t=\int_0^t F_u d B_u^H$, where $\left(F_u, 0 \leq u \leq T\right)$ is a stochastic process in $\mathcal{L}(0, T)$. Assume that there is an $\alpha>1-H$ such that
	$$
	\mathbb{E}\left|F_u-F_v\right|^2 \leq C|u-v|^{2 \alpha},
	$$
	where $|u-v| \leq \delta$ for some $\delta>0$ and
	$$
	\lim _{0 \leq u, v \leq t,|u-v| \rightarrow 0} \mathbb{E}\left|D_u^\phi\left(F_u-F_v\right)\right|^2=0 .
	$$
	Let $f: \mathbb{R}_{+} \times \mathbb{R} \rightarrow \mathbb{R}$ be a function having the first continuous derivative in its first variable and the second continuous derivative in its second variable. Assume that these derivatives are bounded. Moreover, it is assumed that $\mathbb{E} \int_0^T\left|F_s D_s^\phi \eta_s\right| d s<\infty$ and $\left(f^{\prime}\left(s, \eta_s\right) F_s, s \in[0, T]\right)$ is in $\mathcal{L}(0, T)$. Then, for $0 \leq t \leq T$,
	$$
	\begin{aligned}
		f\left(t, \eta_t\right)=& f(0,0)+\int_0^t \frac{\partial f}{\partial s}\left(s, \eta_s\right) d s+\int_0^t \frac{\partial f}{\partial x}\left(s, \eta_s\right) F_s d B_s^H \\
		&+\int_0^t \frac{\partial^2 f}{\partial x^2}\left(s, \eta_s\right) F_s D_s^\phi \eta_s d s. \quad a . s .
	\end{aligned}
	$$
\end{lemma}

\subsection{Fractional Ornstein-Uhlenbeck process}
The fractional Ornstein-Uhlenbeck process
\begin{displaymath}
	Z_{t}^{H}=\int_{-\infty}^{t} e^{-a(t-s)} d B_{s}^{H}=B_{t}^{H}-a \int_{-\infty}^{t} e^{-a(t-s)} B_{s}^{H} d s 
\end{displaymath}
is a zero-mean, stationary Gaussian process, with variance
\begin{displaymath}
	\sigma_{\mathrm{ou}}^{2}=\mathbb{E}\left[\left(Z_{t}^{H}\right)^{2}\right]=\frac{1}{2} a^{-2 H} \Gamma(2 H+1) \sigma_{H}^{2}
\end{displaymath}
and covariance
\begin{displaymath}
	\begin{aligned}
		\mathbb{E}\left[Z_{t}^{H} Z_{t+s}^{H}\right]&=\sigma_{\text {ou }}^{2} \frac{1}{\Gamma(2 H+1)}\left[\frac{1}{2} \int_{\mathbb{R}} e^{-|v|}|a s+v|^{2 H} d v-|a s|^{2 H}\right]\\
		&=\sigma_{\text {ou }}^{2} \frac{2 \sin (\pi H)}{\pi} \int_{0}^{\infty} \cos ( asx ) \frac{x^{1-2 H}}{1+x^{2}} d x .
	\end{aligned}
\end{displaymath}

Note that $Z_{t}^{H}$ is neither a martingale nor a Markov process. The fractional Ornstein-Uhlenbeck process also has the following representation of moving-average integral
\begin{displaymath}
	Z_{t}^{H}=\int_{-\infty}^{t} \mathcal{K}(t-s) d B_{s},
\end{displaymath}
where
\begin{displaymath}
	\mathcal{K}(t)=\frac{1}{\Gamma\left(H+\frac{1}{2}\right)}\left[t^{H-\frac{1}{2}}-a \int_{0}^{t}(t-s)^{H-\frac{1}{2}} e^{-a s} d s\right].
\end{displaymath}
The following lemma introduces the quadratic covariance of Brownian motion and the integral of the fractional  Ornstein-Uhlenbeck process.
\begin{lemma}[\cite{34} Lemma A.1]
	\label{lem2}
		Let 
	\begin{displaymath}
		\psi_{t}=\mathbb{E}\left[\int_{0}^{T} Z_{s}^{H} d s \mid \mathcal{F}_{t}\right].
	\end{displaymath}
	Then $\left(\psi_{t}\right)_{t \in[0, T]}$ is a Gaussian square-integrable martingale and 
	\begin{displaymath}
		d\langle\psi_t, B_t\rangle=\left(\int_{0}^{T-t} \mathcal{K}(s) d s\right) d t=\theta_{t, T} d t,  
	\end{displaymath}
	where 
	\begin{displaymath}
		\theta_{t, T}=\int_{t}^{T} \mathcal{K}(v-t) d v=\int_{0}^{T-t} \mathcal{K}(v) d v .
	\end{displaymath}
\end{lemma}

\section{Main results}
\label{sec:main}
Suppose that $Z_{t}^{H} $ is adapted to the Brownian motion $B'_{t}$. Moreover, $B_{t}$ and $B'_{t}$ are two standard Brownian motions with correlation coefficient $\rho$. In this section we consider an option pricing problem when the dynamics of the underlying asset is driven by the following stochastic differential equation
\begin{equation}       
	\label{new}		
		\left\{\begin{array}{l}
			d X_{t}=\mu X_{t} d t+v_{t}X_{t} d B_{t}, \\
			v_{t}=\bar{v}\left(t \right) +F\left(\gamma Z_{t}^{H} \right) ,
		\end{array}\right.
\end{equation}
where $F$ and $\bar{v}$ are smooth, positive-valued functions, bounded away from zero, with bounded derivatives.

Our objective is to calculate the price of the following derivative
	\begin{equation}
	\label{der}
	W_{t}:=\mathbb{E}\left[g\left(X_{T}\right) \mid \mathcal{F}_{t}\right].
    \end{equation}
For notational simplicity, we introduce the operator
\begin{displaymath}
	\mathcal{L}_{\bar{v}\left ( t \right )} =\partial_{t} +\mu x\partial_{x}+\frac{1}{2} \bar{v}\left ( t \right )^{2}x^{2}\partial^{2}_{xx}.
\end{displaymath}

The following Theorem \ref{thm:5} gives an approximate expression of derivative price.
\begin{theorem}\label{thm:5}
	If the underlying asset and the derivative follow the dynamics given by (\ref{new}) (\ref{der}), we approximate the price of the derivative as follows
	\begin{displaymath}
		W_{t}=M\left(t,X_{t} \right)+ O\left(\gamma^{2} \right),
	\end{displaymath}
	where
	\begin{displaymath}
		\begin{aligned}
			M\left(t,X_{t} \right)=&M_{1} \left ( t,X_{t} \right )\\
			&+a\left(t,X_{t} \right)\gamma\bar{v}\left ( t \right )\phi_{t}\left(x^{2}\partial_{xx}^{2} \right)M_{1}\left(t,X_{t} \right)+a\left(t,X_{t} \right)\gamma \rho M_{2} \left ( t,X_{t} \right )\\
			&+\gamma\phi_{t}M_{3} \left ( t,X_{t} \right ) +\gamma M_{4} \left ( t,X_{t} \right )+\gamma \rho M_{5} \left ( t,X_{t} \right ),
		\end{aligned}
	\end{displaymath}
	$\phi_{t}=\mathbb{E}\left[\int_{t}^{T} Z_{s}^{H} d s \mid \mathcal{F}_{t}\right]$ and $M_{1} \left ( t,x \right ), M_{2} \left ( t,x \right ), M_{3} \left ( t,x \right ), M_{4} \left ( t,x \right ), M_{5} \left ( t,x \right ), a\left(t,x \right)$ are deterministic and can be solved by the following equations,
	\begin{equation}
		\label{equ}
		\left\{
		\begin{aligned}
			&\mathcal{L}_{\bar{v}\left ( t \right )}M_{1} \left ( t,x \right ) =0 ,\\
			&\mathcal{L}_{\bar{v}\left ( t \right )}M_{2} \left ( t,x \right )=-\bar{v}\left ( t \right )^{2} \left (x \partial_{x}(x^{2}\partial_{xx}^{2} ) \right )  M_{1}\left ( t,x \right )\theta _{t,T},\\
			&\mathcal{L}_{\bar{v}\left ( t \right )}M_{3} \left ( t,x \right ) =-\left(x^{2}\partial_{xx}^{2} \right)M_{1}\left(t,x \right)\left[a\left ( t,x \right ){\bar{v}}' \left ( t \right )+\bar{v} \left ( t \right )\mathcal{L}_{\bar{v}\left ( t \right )}a \left ( t,x \right ) \right] ,\\
			&\mathcal{L}_{\bar{v}\left ( t \right )}M_{4} \left ( t,x \right )=- \bar{v}\left(t \right) \left( x\partial_{x}\right) M_{3} \left ( t,x \right )\theta_{t,T}  ,\\
			&\mathcal{L}_{\bar{v}\left ( t \right )}M_{5} \left ( t,x \right )=- M_{2} \left ( t,x \right ) \mathcal{L}_{\bar{v}\left ( t \right )}a \left ( t,x \right ) ,\\
			&\left ( 1-a\left ( t,x \right )  \right )\bar{v}\left ( t \right )  \left( x^{2}\partial^{2}_{xx}\right) M_{1} \left ( t,x \right )-M_{3} \left ( t,x \right )=0,\\
			&M_{1} \left ( T,x \right )=g(x),\\
			&M_{2} \left ( T,x \right )=M_{3} \left ( T,x \right )=M_{4} \left ( T,x \right )=M_{5} \left ( T,x \right )=0.
		\end{aligned}
		\right.
	\end{equation}

	\begin{proof}
		For smooth function $M_{1} \left ( t,x \right )$, we have by Itô's formula
		\begin{displaymath}
			\begin{aligned}
			   dM_{1} \left ( t,X_{t} \right )=&\left(\bar{v}\left ( t \right )F\left( \gamma Z_{t}^{H}\right) +\frac{F\left( \gamma Z_{t}^{H}\right)^{2}}{2} \right) \left( x^{2}\partial^{2}_{xx}\right)  M_{1}\left ( t,X_{t} \right )dt+v_{t} \left( x\partial_{x}\right) M_{1} \left ( t,X_{t} \right )dB_{t}\\
			   =&\left(\gamma\bar{v}\left ( t \right ) Z_{t}^{H}+\frac{\gamma^{2} g^{\gamma}\left(Z_{t}^{H} \right) }{2} \right) \left( x^{2}\partial^{2}_{xx}\right)  M_{1}\left ( t,X_{t} \right )dt+v_{t} \left( x\partial_{x}\right) M_{1} \left ( t,X_{t} \right )dB_{t},
		    \end{aligned}
		\end{displaymath}
		\begin{displaymath}
			\begin{aligned}
				d\left(\phi_{t}\left(x^{2}\partial_{xx}^{2} \right)M_{1}\left(t,X_{t} \right) \right) =&	\left(x^{2}\partial_{xx}^{2} \right)M_{1}\left(t,X_{t} \right)d\phi_{t}+\phi_{t}d\left[\left(x^{2}\partial_{xx}^{2} \right)M_{1}\left(t,X_{t} \right) \right]\\
				=&\left(x^{2}\partial_{xx}^{2} \right)M_{1}\left(t,X_{t} \right)d\phi_{t}\\
				&+\phi_{t}  \left ( \partial_{t}(x^{2}\partial_{xx}^{2} ) \right )M_{1}\left ( t,X_{t} \right )  dt \\
				&+\phi_{t}  \left ( \partial_{x}(x^{2}\partial_{xx}^{2} ) \right )M_{1}\left ( t,X_{t} \right )  dX_{t}\\
				&+\frac{1}{2}\phi_{t}  \left ( \partial^{2}_{xx}(x^{2}\partial_{xx}^{2} ) \right )M_{1}\left ( t,X_{t} \right )  d\left \langle X_{t},X_{t} \right \rangle \\
				&+\left ( \partial_{x}(x^{2}\partial_{xx}^{2} ) \right )  M_{1}\left ( t,X_{t} \right ) d\left \langle \phi_{t},X_{t} \right \rangle\\
				=&-Z_{t}^{H}\left(x^{2}\partial_{xx}^{2} \right)M_{1}\left(t,X_{t} \right)dt+\left(x^{2}\partial_{xx}^{2} \right)M_{1}\left(t,X_{t} \right)d\psi_{t}\\
				&+\phi_{t}v_{t}  \left (x \partial_{x}(x^{2}\partial_{xx}^{2} ) \right )M_{1}\left ( t,X_{t} \right )  dB_{t}\\
				&+\phi_{t}\left(\gamma\bar{v}\left ( t \right ) Z_{t}^{H}+\frac{\gamma^{2} g^{\gamma}\left(Z_{t}^{H} \right) }{2} \right) \left (x^{2} \partial^{2}_{xx}(x^{2}\partial_{xx}^{2} ) \right )M_{1}\left ( t,X_{t} \right )  dt \\
				&+v_{t}\left (x \partial_{x}(x^{2}\partial_{xx}^{2} ) \right )  M_{1}\left ( t,X_{t} \right ) d\left \langle \phi_{t},B_{t} \right \rangle,
			\end{aligned}
		\end{displaymath}
		where 
		$$
		g^{\gamma}\left(y \right)=2\bar{v}\left ( t \right )\frac{F\left( \gamma y\right) -\gamma y}{\gamma^{2}}+\frac{F\left( \gamma y\right)^{2}}{\gamma^{2}}.$$ 
		We have $\left\langle\phi_t, B_t\right\rangle=\rho\langle\psi, B'_t\rangle$, and therefore, 
		\begin{displaymath}
			\begin{aligned}
				&d\left(M_{1} \left ( t,X_{t} \right )+a\left(t,X_{t} \right) \gamma\bar{v}\left ( t \right )\phi_{t}\left(x^{2}\partial_{xx}^{2} \right)M_{1}\left(t,X_{t} \right) \right) \\
				=&\left ( \gamma \left ( 1-a\left ( t,X_{t} \right )  \right )\bar{v}\left ( t \right )Z_{t}^{H} + \frac{\gamma^{2} g^{\gamma}\left(Z_{t}^{H} \right) }{2}  \right )  \left( x^{2}\partial^{2}_{xx}\right) M_{1} \left ( t,X_{t} \right )dt\\
				&+a\left(t,X_{t} \right)\phi_{t}\left(\gamma^{2}\bar{v}\left ( t \right )^{2} Z_{t}^{H}+\frac{\gamma^{3}\bar{v}\left ( t \right ) g^{\gamma}\left(Z_{t}^{H} \right)}{2} \right)  \left (x^{2} \partial^{2}_{xx}(x^{2}\partial_{xx}^{2} ) \right )M_{1}\left ( t,X_{t} \right )  dt \\
				&+a\left(t,X_{t} \right)\gamma\rho\bar{v}\left ( t \right )v_{t}X_{t}\left ( \partial_{x}(x^{2}\partial_{xx}^{2} ) \right )  M_{1}\left ( t,X_{t} \right ) d\left \langle \psi_{t},B'_{t} \right \rangle\\
				&+\gamma\left(\phi_{t}\left(x^{2}\partial_{xx}^{2} \right)M_{1}\left(t,X_{t} \right) \right)da\left(t,X_{t} \right)\bar{v}\left ( t \right )+M_{t}^{\left(1 \right) },
			\end{aligned}
		\end{displaymath}
		where $M_{t}^{\left(1 \right) }$ is a martingale satisfying
		\begin{displaymath}
			\begin{aligned}
				dM_{t}^{\left(1 \right) }=&a\left(t,X_{t} \right) v_{t} X_{t}\partial_{x}M_{1} \left ( t,X_{t} \right )dB_{t}\\
				&+a\left(t,X_{t} \right)\gamma\bar{v}\left ( t \right )\left(x^{2}\partial_{xx}^{2} \right)M_{1}\left(t,X_{t} \right)d\psi_{t}\\
				&+a\left(t,X_{t} \right)\gamma\bar{v}\left ( t \right )\phi_{t}v_{t}X_{t}  \left ( \partial_{x}(x^{2}\partial_{xx}^{2} ) \right )M_{1}\left ( t,X_{t} \right )  dB_{t}.
			\end{aligned}
		\end{displaymath}
		Applying Lemma \ref{lem2}, notice that $d\langle\psi_t, B'_t\rangle=\theta_{t, T} d t$, thus we can write 	
		\begin{displaymath}
			\begin{aligned}
				&d\left(M_{1} \left ( t,X_{t} \right )+a\left(t,X_{t} \right)\gamma\bar{v}\left ( t \right )\phi_{t}\left(x^{2}\partial_{xx}^{2} \right)M_{1}\left(t,X_{t} \right)\right) \\
				&+d\left( a\left(t,X_{t} \right)\gamma \rho M_{2} \left ( t,X_{t} \right )+\gamma\phi_{t}M_{3} \left ( t,X_{t} \right )  \right) \\
				=&\gamma \rho M_{2} \left ( t,X_{t} \right )da\left(t,X_{t} \right)+\gamma v_{t}\left( x\partial_{x}\right) M_{3} \left ( t,X_{t} \right )\theta_{t, T} d t \\
				&+dR_{t}^{\left( 1\right) }+dM_{t}^{\left(1 \right) }+dM_{t}^{\left(2 \right) },\\
			\end{aligned}
		\end{displaymath}
		where
		\begin{displaymath}
			\begin{aligned}
				dR_{t}^{\left( 1\right) }=&\frac{\gamma^{2} g^{\gamma}\left(Z_{t}^{H} \right)}{2}  \left( x^{2}\partial^{2}_{xx}\right) M_{1} \left ( t,X_{t} \right )dt\\
				&+a\left ( t,X_{t} \right )\phi_{t}\left(\gamma^{2}\bar{v}\left ( t \right )^{2} Z_{t}^{H}+\frac{\gamma^{3}\bar{v}\left ( t \right ) g^{\gamma}\left(Z_{t}^{H} \right)}{2} \right)  \left (x^{2} \partial^{2}_{xx}(x^{2}\partial_{xx}^{2} ) \right )M_{1}\left ( t,X_{t} \right )  dt \\
				&+a\left ( t,X_{t} \right )\gamma^{2}\rho\bar{v}\left ( t \right )Z_{t}^{H}\left (x \partial_{x}(x^{2}\partial_{xx}^{2} ) \right )  M_{1}\left ( t,X_{t} \right ) \theta_{t, T} d t \\
				&+\left(a\left ( t,X_{t} \right )\gamma^{2} \rho \bar{v}\left ( t \right ) Z_{t}^{H}+\frac{a\left ( t,X_{t} \right )\gamma^{3} \rho g^{\gamma}\left(Z_{t}^{H} \right)}{2} \right) \left(x^{2} \partial^{2}_{xx}\right) M_{2} \left ( t,X_{t} \right )dt\\
				&+\frac{1}{2}\left(\phi_{t}\left(x^{2}\partial_{xx}^{2} \right)M_{1}\left(t,X_{t} \right) \right)\bar{v} \left ( t \right )\left(\gamma^{2}\bar{v}\left ( t \right ) Z_{t}^{H}+\frac{\gamma^{3} g^{\gamma}\left(Z_{t}^{H} \right)}{2} \right)\left(x^{2} \partial^{2}_{xx}\right) a\left ( t,X_{t} \right )dt\\
				&+\phi_{t}\left(\gamma^{2}\bar{v}\left ( t \right ) Z_{t}^{H}+\frac{\gamma^{3} g^{\gamma}\left(Z_{t}^{H} \right)}{2} \right) \left( x^{2}\partial^{2}_{xx}\right) M_{3} \left ( t,X_{t} \right )dt,\\    			
			\end{aligned}
		\end{displaymath}
	\begin{displaymath}
		\begin{aligned}   			
			dM_{t}^{\left( 2\right) }=&\gamma\left(\phi_{t}\left(x^{2}\partial_{xx}^{2} \right)M_{1}\left(t,X_{t} \right) \right)\bar{v} \left ( t \right )v_{t}\left( x\partial_{x}\right) a\left ( t,X_{t} \right )dB_{t}\\
			&+a\left ( t,X_{t} \right )\gamma \rho v_{t} \left( x\partial_{x}\right) M_{2} \left ( t,X_{t} \right )dB_{t}\\
			&+\gamma M_{3} \left ( t,X_{t} \right )d\psi_{t}\\
			&+\gamma v_{t}\phi_{t}(x\partial_{x})M_{3} \left ( t,X_{t} \right )dB_{t}.
		\end{aligned}
	\end{displaymath}
		In summary, we have
		\begin{displaymath}
			\begin{aligned}
				&d\left(M_{1} \left ( t,X_{t} \right )+a\left(t,X_{t} \right)\gamma\bar{v}\left ( t \right )\phi_{t}\left(x^{2}\partial_{xx}^{2} \right)M_{1}\left(t,X_{t} \right)\right) \\
				&+d\left( a\left(t,X_{t} \right)\gamma \rho M_{2} \left ( t,X_{t} \right )+\gamma\phi_{t}M_{3} \left ( t,X_{t} \right )\right)  \\
				&+d\left( \gamma M_{4} \left ( t,X_{t} \right )+\gamma \rho M_{5} \left ( t,X_{t} \right ) \right) \\
				=&dR_{t}^{\left( 1\right) }+dR_{t}^{\left( 2\right) }+dM_{t}^{\left(1 \right) }+dM_{t}^{\left(2 \right) }+dM_{t}^{\left(3 \right) },
			\end{aligned}
		\end{displaymath}
		where
		\begin{displaymath}
			\begin{aligned}
				dR_{t}^{\left( 2\right) }=&M_{2} \left ( t,X_{t} \right )\left(\gamma^{2} \rho \bar{v}\left ( t \right ) Z_{t}^{H}+\frac{\gamma^{3} \rho g^{\gamma}\left(Z_{t}^{H} \right)}{2} \right)dt\\
				&+\gamma^{2}\rho Z_{t}^{H}\left( x\partial_{x}\right) M_{3} \left ( t,X_{t} \right )\theta_{t, T} d t \\
				&+\left(\gamma^{2}\bar{v}\left ( t \right ) Z_{t}^{H}+\frac{\gamma^{3} g^{\gamma}\left(Z_{t}^{H} \right)}{2} \right) \left( x^{2}\partial^{2}_{xx}\right) M_{4} \left ( t,X_{t} \right )dt\\
				&+\left(\gamma^{2} \rho\bar{v}\left ( t \right ) Z_{t}^{H}+\frac{\gamma^{3}\rho g^{\gamma}\left(Z_{t}^{H} \right)}{2} \right) \left( x^{2}\partial^{2}_{xx}\right) M_{5} \left ( t,X_{t} \right )dt,\\ 
				dM_{t}^{\left(3 \right) }=&v_{t}X_{t}M_{2} \left ( t,X_{t} \right )dB_{t}+\gamma v_{t} (x\partial_{x})M_{4} \left ( t,X_{t} \right )dB_{t}+\gamma\rho v_{t} (x\partial_{x})M_{5} \left ( t,X_{t} \right )dB_{t}.  	
			\end{aligned}
		\end{displaymath}
		Therefore,
		\begin{displaymath}
			\begin{aligned}    			
				dM\left(t,X_{t} \right)=dM_{t}^{\left(1 \right) }+dM_{t}^{\left(2 \right) }+dM_{t}^{\left(3 \right) }+dR_{t}^{\left( 1\right) }+dR_{t}^{\left( 2\right) },
			\end{aligned}
		\end{displaymath}
		where $M_{t}^{\left(1 \right) }$, $M_{t}^{\left(2 \right) }$, $M_{t}^{\left(3 \right) }$ are martingales. Let 
		\begin{displaymath}        		
			M_{t}=M_{t}^{\left(1 \right) }+M_{t}^{\left(2 \right) }+M_{t}^{\left(3 \right) },\ R_{t}=R_{t}^{\left( 1\right) }+R_{t}^{\left( 2\right) }.
		\end{displaymath}
		In addition, according to (\ref{equ}), we have $M\left(T,X_{T} \right)=g\left(X_{T} \right) $ and
		\begin{displaymath}
			\begin{aligned}
				W_{t}=&\mathbb{E}\left[g\left(X_{T}\right) \mid \mathcal{F}_{t}\right]\\
				=&\mathbb{E}\left[M \left ( T,X_{T} \right ) \mid \mathcal{F}_{t}\right]\\
				=&M \left ( t,X_{t} \right )+\mathbb{E}\left[ M_{T}-M_{t} \mid \mathcal{F}_{t} \right]+ \mathbb{E}\left[ \int_{t}^{T}  dR_{s} \mid \mathcal{F}_{t} \right]\\
				=&M \left ( t,X_{t} \right )+ \mathbb{E}\left[ \int_{t}^{T}  dR_{s} \mid \mathcal{F}_{t} \right].
			\end{aligned}
		\end{displaymath}
	    Note that $g^\gamma(y)$ is bounded uniformly in $\gamma$ by
     	$$
	    \left|g^\gamma(y)\right| \leq\left(\left\|\bar{v}\right\|_{\infty}\left\|F^{\prime \prime}\right\|_{\infty}+\left\|F^{\prime}\right\|_{\infty}^2\right) y^2 .
     	$$
		This completes the proof of Theorem \ref{thm:5} since $\mathbb{E}\left[ \int_{t}^{T}  dR_{s} \mid \mathcal{F}_{t} \right]$ is of order $\gamma^{2}$.
	\end{proof}
\end{theorem}

\section{Application: European option pricing under the fractional Stein-Stein volatility model}
\label{example}
In this subsection we calculate the approximate pricing of a European call option with an execution price of $K$ under the fractional Stein-Stein volatility model as an example, i.e.
\begin{equation}
	\label{equ01}
	\left\{\begin{array}{l}
		d X_{t}=\mu X_{t} d t+v_{t}X_{t} d B_{t}, \\
		d v_{t}=\beta\left(\alpha-v_{t}\right) d t+\gamma d B_{t}^{H},
	\end{array}\right.
\end{equation}
\begin{equation}
	\label{wt}
W_{t}:=\mathbb{E}\left[\left(S_{T}-K\right)^{+} \mid \mathcal{F}_{t}\right],
\end{equation}
where $X_{t}$ is risky asset price process. Here, $\mu$ is the drift rate of the risk asset price process and $\beta$ is the average recurrent rate of the volatility process. Since the volatility process is a mean-reverting process, $v_{t}$ tends towards a long-term value $\alpha$ with rate $\beta$. $\gamma$  is a constant. $ B_{t}^{H}$ is a fractional Brownian motion with Hurst parameter $H>1 / 2$.

Based on Lemma \ref{thm:1}, we have the following Lemma \ref{lem:1}.

\begin{lemma}\label{lem:1}
	The equation  
	\begin{equation}
		\label{equ0}
		dv(t)=\beta \left( \alpha- v(t)\right) dt+\gamma dB_{t}^{H}
	\end{equation}	
	has a unique solution of the form
	\begin{displaymath}
		\begin{aligned}
			v_{t}=&e^{-\beta t} v_{0}+\beta \alpha \int_{0}^{t} e^{\beta(s-t)} d s+\gamma B_{t}^{H}-\beta \int_{0}^{t} \gamma e^{\beta(s-t)} B_{s}^{H} d  s\\
			=&\bar{v}\left(t \right) +\gamma Z_{t}^{H}.
		\end{aligned}		
	\end{displaymath}
\end{lemma}

If the underlying asset and the derivative follow the dynamics given by equations (\ref{equ01}) (\ref{wt}), applying the Theorem \ref{thm:5}, our approximate pricing method proceeds as follows.

Step 1: We solve $M_{1} \left ( t,x \right )$ by
\begin{equation}
	\label{pde1}
	\left\{\begin{array}{l}
		\mathcal{L}_{\bar{v}\left ( t \right )}M_{1} \left ( t,x \right ) =0, \\
		M_{1} \left ( T,x \right )=\left(x-K\right)^{+}.
	\end{array}\right.
\end{equation}  
Let 
\begin{displaymath} 
	\left\{\begin{array}{l}
		u=M_{1}\left(t,x \right) ,\\
		y=xe^{\mu\left(T-t \right) },
	\end{array}\right.
\end{displaymath}
equation (\ref{pde1}) becomes
\begin{equation}
	\label{pde2}
	\left\{\begin{array}{l}
		\partial_{t}u  +\frac{1}{2} \bar{v}\left ( t \right )^{2}y^{2}\partial^{2}_{yy}u  =0 ,\\
		u \mid _{t=T}=\left(y-K\right)^{+}.
	\end{array}\right.
\end{equation} 
Then let $\tau =\int_{0}^{t} \bar{v}\left ( s \right )^{2} ds$ , equation (\ref{pde2}) becomes
\begin{displaymath}
	\left\{\begin{array}{l}
		\partial_{\tau}u +\frac{1}{2} y^{2}\partial^{2}_{yy}u=0, \\
		u \mid _{\tau =\hat{T}}=\left(y-K\right)^{+},
	\end{array}\right.
\end{displaymath}
where $\hat{T}=\int_{0}^{T} \bar{v}\left ( s \right )^{2} ds$.
Applying the Black-Scholes formula, we get 
\begin{displaymath}
	M_{1}\left(t,x \right)=u(y,\tau)=yN(\hat{d_{1}})-KN(\hat{d_{2}})=xe^{\mu\left(T-t \right) }N(\hat{d_{1}})-KN(\hat{d_{2}}),
\end{displaymath}
where
\begin{displaymath}
	\hat{d_{1}}=\frac{\ln{\frac{y}{K} }+\frac{1}{2} \left ( \hat{T}-\tau  \right )  }{\sqrt{ \hat{T}-\tau } } =\frac{\ln{\frac{x}{K} }+\mu \left ( T-t \right ) +\frac{1}{2} \int_{t}^{T}\bar{v}\left ( s \right )^{2}ds  }{\sqrt{ \int_{t}^{T}\bar{v}\left ( s \right )^{2}ds  } },
\end{displaymath}
\begin{displaymath}
	\hat{d_{2}}=\hat{d_{1}}-\sqrt{ \hat{T}-\tau }=\hat{d_{1}}-\sqrt{ \int_{t}^{T}\bar{v}\left ( s \right )^{2}ds  } ,
\end{displaymath}
\begin{displaymath}
	N\left(x \right)=\frac{1}{\sqrt{2\pi } }\int_{-\infty }^{x}  e^{-\frac{s^{2}}{2} }ds .
\end{displaymath}
Moreover,  we have  
\begin{displaymath}
	\partial_{x}M_{1}\left(t,x \right)=e^{\mu\left(T-t \right) }N(\hat{d_{1}})+\frac{e^{\mu\left(T-t \right) }e^{-\frac{\hat{d_{1}}^{2}}{2} }}{\sqrt{2\pi \int_{t}^{T}\bar{v}\left ( s \right )^{2}ds  }}-\frac{Ke^{-\frac{\hat{d_{2}}^{2}}{2} }}{x\sqrt{2\pi \int_{t}^{T}\bar{v}\left ( s \right )^{2}ds  }} ,
\end{displaymath}
\begin{displaymath}
	\begin{aligned}
		\partial^{2}_{xx}M_{1}\left(t,x \right)=&\frac{e^{\mu\left(T-t \right) }e^{-\frac{\hat{d_{1}}^{2}}{2} }}{x\sqrt{2\pi \int_{t}^{T}\bar{v}\left ( s \right )^{2}ds  }}-\frac{e^{\mu\left(T-t \right) }e^{-\frac{\hat{d_{1}}^{2}}{2} }\hat{d_{1}}}{x\sqrt{2\pi }\int_{t}^{T}\bar{v}\left ( s \right )^{2}ds } \\
		&+\frac{Ke^{-\frac{\hat{d_{2}}^{2}}{2} }\hat{d_{2}}}{x^{2}\sqrt{2\pi }\int_{t}^{T}\bar{v}\left ( s \right )^{2}ds } +\frac{Ke^{-\frac{\hat{d_{2}}^{2}}{2} }}{x^{2}\sqrt{2\pi \int_{t}^{T}\bar{v}\left ( s \right )^{2}ds} } .
	\end{aligned}
\end{displaymath}

Step 2: We solve $M_{2} \left ( t,x \right )$ by
\begin{equation}
	\label{pde3}
	\left\{\begin{array}{l}
		\mathcal{L}_{\bar{v}\left ( t \right )}M_{2} \left ( t,x \right )  =-\bar{v}\left ( t \right )^{2} \left (x \partial_{x}(x^{2}\partial_{xx}^{2} ) \right )  M_{1}\left ( t,x \right )\theta _{t,T} =\overline{M_{1}}\left ( t,x \right ) ,  \\
		M_{2} \left ( T,x \right )=0.
	\end{array}\right.
\end{equation}
Let 
\begin{displaymath}
	\left\{\begin{array}{l}
		z=\text{ln}x, \\
		\tau=T-t,
	\end{array}\right.
\end{displaymath}
equation (\ref{pde3}) becomes
\begin{equation}
	\label{pde4}
	\left\{\begin{array}{l}
		\partial_{\tau}M_{2}  -\frac{\bar{v}^{2}}{2} \partial^{2}_{zz}M_{2}-\left( \mu-\frac{\bar{v}^{2}}{2}  \right)\partial_{z}M_{2} =\overline{M_{1}}, \\
		M_{2} \mid _{\tau=0}=0.
	\end{array}\right.
\end{equation}
Let
\begin{displaymath}
	M_{2}=ue^{\alpha \tau +\beta z},
\end{displaymath}
\begin{displaymath}
	\zeta=\int_{0}^{\tau} \bar{v}^{2} ds,
\end{displaymath}
where
\begin{displaymath}
	\alpha=\left(\frac{1}{2}+\frac{\mu}{\bar{v}^{2}} \right) \left(\frac{3\mu}{2}-\frac{\bar{v}^{2}}{4} \right) ,
\end{displaymath}
\begin{displaymath}
	\beta=\frac{1}{2}+\frac{\mu}{\bar{v}^{2}}.
\end{displaymath}
Then equation (\ref{pde4}) becomes
\begin{equation}
	\label{pde5}
	\left\{\begin{array}{l}
		\partial_{\zeta}u  - \frac{1}{2}\partial^{2}_{zz}u=\widetilde {M_{1}}\left(\zeta,z \right),  \\
		u \mid _{\zeta=0}=0,
	\end{array}\right.
\end{equation}
where
\begin{displaymath}
	\widetilde {M_{1}}\left(\zeta,z \right)=\frac{\overline{M_{1}}}{e^{\alpha \tau +\beta z}\bar{v}^{2}}.
\end{displaymath}
We get the solution of equation (\ref{pde5}) as follows
\begin{displaymath}
	u\left(\zeta,z \right) =\int_{0}^{\zeta}\int_{z-\frac{1}{2}\left(\zeta-m \right) }^{z+\frac{1}{2}\left(\zeta-m \right)}\widetilde {M_{1}}\left(m,n \right)dndm,
\end{displaymath}
\begin{displaymath}
	M_{2}=e^{\alpha \tau +\beta z}\int_{0}^{\zeta}\int_{z-\frac{1}{2}\left(\zeta-m \right) }^{z+\frac{1}{2}\left(\zeta-m \right)}\widetilde {M_{1}}\left(m,n \right)dndm.
\end{displaymath}

Step 3: We solve $M_{3} \left ( t,x \right )$ and $a(t,x)$ by
\begin{equation}
	\label{pde6}
	\left\{\begin{array}{l}
		\mathcal{L}_{\bar{v}\left ( t \right )}M_{3} \left ( t,x \right ) =-\left(x^{2}\partial_{xx}^{2} \right)M_{1}\left(t,x \right)\left[a\left ( t,x \right ){\bar{v}}' \left ( t \right )+\bar{v} \left ( t \right )\mathcal{L}_{\bar{v}\left ( t \right )}a \left ( t,x \right ) \right] , \\
		\left ( 1-a\left ( t,x \right )  \right )\bar{v}\left ( t \right )  \left( x^{2}\partial^{2}_{xx}\right) M_{1} \left ( t,x \right )-M_{3} \left ( t,x \right )=0,\\
		M_{3} \left ( T,x \right )=0.
	\end{array}\right.
\end{equation}
Let $f\left(t,x \right)=\left(  x^{2}\partial_{xx}^{2} \right)M_{1}\left(t,x \right)$. Then we have
\begin{displaymath}
	\begin{aligned}
		\partial_{t}M_{3}\left(t,x \right) =&{\bar{v}}' \left ( t \right )\left [ 1-a\left ( t,x \right )  \right ] f\left ( t,x \right )  \\
		&+\bar{v}\left ( t \right )\left [ 1-a\left ( t,x \right )  \right ] \partial_{t}f\left ( t,x \right ) -\bar{v}\left ( t \right ) f\left ( t,x \right )\partial_{t}a\left ( t,x \right ),\\
		\partial_{x}M_{3}\left(t,x \right) =&\bar{v}\left ( t \right )\left [ 1-a\left ( t,x \right )  \right ] \partial_{x}f\left ( t,x \right )\\
		&-\bar{v}\left ( t \right ) f\left ( t,x \right )\partial_{x}a\left ( t,x \right ),\\
		\partial^{2}_{xx}M_{3}\left(t,x \right) =&\bar{v}\left ( t \right )\left [ 1-a\left ( t,x \right )  \right ] \partial^{2}_{xx}f\left ( t,x \right )\\
		&-2\bar{v}\left ( t \right ) \partial_{x}f\left ( t,x \right ) \partial_{x}a\left ( t,x \right )-\bar{v}\left ( t \right )f\left ( t,x \right )\partial^{2}_{xx}a\left(t,x \right).
	\end{aligned}
\end{displaymath} 
Let
\begin{displaymath}
	\tilde{a}\left(t,x \right) =a\left(t,x \right)-1,z=\text{ln}x,
\end{displaymath} 
equation (\ref{pde6}) is translated into
\begin{equation}
	\label{pde7}
	\left\{\begin{array}{l}
		\left[\bar{v}\left ( t \right )\partial_{t}f\left ( t,x \right )+\mu \bar{v}\left ( t \right )\partial_{z}f\left ( t,x \right )+\frac{1}{2}\bar{v}\left ( t \right )^{3}\partial^{2}_{zz}f\left ( t,x \right )-\frac{1}{2}\bar{v}\left ( t \right )^{3}\partial_{z}f\left ( t,x \right ) \right]\tilde{a}\left(t,x \right)\\
		+\bar{v}\left ( t \right )^{3}\partial_{z}f\left ( t,x \right ) \partial_{z}\tilde{a}\left ( t,x \right )={\bar{v}}' \left ( t \right )f\left ( t,x \right ) ,\\
		M_{3} \left ( t,x \right )=\tilde{a}\left(t,x \right)\bar{v}\left ( t \right )  f\left ( t,x \right ),\\
		\tilde{a}\left(T,x \right)=0.
	\end{array}\right.
\end{equation} 
The solutions of equation (\ref{pde7}) are as follows
\begin{displaymath}
	\tilde{a}=e^{-\int_{0}^{z}m\left ( t,s \right )ds  }\int_{t}^{T} \int_{0}^{z} n\left ( \tau ,s \right ) dsd\tau ,\ M_{3} \left ( t,x \right )=\tilde{a}\bar{v}\left ( t \right )  f\left ( t,x \right ),
\end{displaymath}
where
\begin{displaymath}
	m\left(t,z \right) =\frac{\partial_{t}f}{\bar{v}^{2}\partial_{z}f}+\frac{\mu}{\bar{v}^{2}}+\frac{\partial^{2}_{zz}f}{2\partial_{z}f}-\frac{1}{2},n\left(t,z \right)=-\partial_{t} \left ( q\left ( t,z \right ) e^{\int_{0}^{z}m\left ( t,s \right )ds  } \right )  ,
\end{displaymath}
\begin{displaymath}
	q\left(t,z \right)=\frac{{\bar{v}}'f}{\bar{v}^{3}\partial_{z}f}.
\end{displaymath}

Step 4: We solve the $M_{4} \left ( t,x \right )$ and $M_{5} \left ( t,x \right )$ by
\begin{equation}
	\label{pde8}
	\left\{\begin{array}{l}
		\mathcal{L}_{\bar{v}\left ( t \right )}M_{4} \left ( t,x \right ) =- \bar{v}\left(t \right) \left( x\partial_{x}\right) M_{3} \left ( t,x \right )\theta_{t,T} =\overline{M_{3}}, \\
		\mathcal{L}_{\bar{v}\left ( t \right )}M_{5} \left ( t,x \right )=- M_{2} \left ( t,x \right ) \mathcal{L}_{\bar{v}\left ( t \right )}a \left ( t,x \right ) =\overline{M_{2}},\\
		M_{4} \left ( T,x \right )=M_{5} \left ( T,x \right )=0.
	\end{array}\right.
\end{equation}
Equation (\ref{pde8}) is solved similarly to $M_{2} \left ( t,x \right )$. Let
\begin{displaymath}
	\left\{\begin{array}{l}
		z=\text{ln}x, \\
		\tau=T-t,
	\end{array}\right.
\end{displaymath}
equation (\ref{pde8}) becomes
\begin{equation}
	\label{pde9}
	\left\{\begin{array}{l}
		\partial_{\tau}M_{4}  -\frac{\bar{v}^{2}}{2} \partial^{2}_{zz}M_{4}-\left( \mu-\frac{\bar{v}^{2}}{2}  \right)\partial_{z}M_{4} =\overline{M_{3}}, \\
		\partial_{\tau}M_{5}  -\frac{\bar{v}^{2}}{2} \partial^{2}_{zz}M_{5}-\left( \mu-\frac{\bar{v}^{2}}{2}  \right)\partial_{z}M_{5} =\overline{M_{2}}, \\
		M_{4} \mid _{\tau=0}=M_{5} \mid _{\tau=0}=0.
	\end{array}\right.
\end{equation}
Let
\begin{displaymath}
	M_{4}=ue^{\alpha \tau +\beta z},M_{5}=\hat{u}e^{\alpha \tau +\beta z},\zeta=\int_{0}^{\tau} \bar{v}^{2} ds,
\end{displaymath}
where
\begin{displaymath}
	\alpha=\left(\frac{1}{2}+\frac{\mu}{\bar{v}^{2}} \right) \left(\frac{3\mu}{2}-\frac{\bar{v}^{2}}{4} \right),\beta=\frac{1}{2}+\frac{\mu}{\bar{v}^{2}}.
\end{displaymath}
Equation (\ref{pde9}) becomes
\begin{displaymath}
	\left\{\begin{array}{l}
		\partial_{\zeta}u  - \frac{1}{2}\partial^{2}_{zz}u=\widetilde {M_{3}}\left(\zeta,z \right) , \\
		\partial_{\zeta}\hat{u}  - \frac{1}{2}\partial^{2}_{zz}\hat{u}=\widetilde {M_{2}}\left(\zeta,z \right),  \\
		u \mid _{\zeta=0}=u \mid _{\zeta=0}=0,
	\end{array}\right.
\end{displaymath}
where
\begin{displaymath}
	\widetilde {M_{3}}\left(\zeta,z \right)=\frac{\overline{M_{3}}}{e^{\alpha \tau +\beta z}\bar{v}^{2}}, \ \widetilde {M_{2}}\left(\zeta,z \right)=\frac{\overline{M_{2}}}{e^{\alpha \tau +\beta z}\bar{v}^{2}}.
\end{displaymath}
We get the solution of the above equation as follows
\begin{displaymath}
	u\left(\zeta,z \right) =\int_{0}^{\zeta}\int_{z-\frac{1}{2}\left(\zeta-m \right) }^{z+\frac{1}{2}\left(\zeta-m \right)}\widetilde {M_{3}}\left(m,n \right)dndm,
\end{displaymath}
\begin{displaymath}
	\hat{u}\left(\zeta,z \right) =\int_{0}^{\zeta}\int_{z-\frac{1}{2}\left(\zeta-m \right) }^{z+\frac{1}{2}\left(\zeta-m \right)}\widetilde {M_{2}}\left(m,n \right)dndm,
\end{displaymath}
\begin{displaymath}
	M_{4}=e^{\alpha \tau +\beta z}\int_{0}^{\zeta}\int_{z-\frac{1}{2}\left(\zeta-m \right) }^{z+\frac{1}{2}\left(\zeta-m \right)}\widetilde {M_{3}}\left(m,n \right)dndm,
\end{displaymath}
\begin{displaymath}
	M_{5}=e^{\alpha \tau +\beta z}\int_{0}^{\zeta}\int_{z-\frac{1}{2}\left(\zeta-m \right) }^{z+\frac{1}{2}\left(\zeta-m \right)}\widetilde {M_{2}}\left(m,n \right)dndm.
\end{displaymath}

\section{Numerical simulations}
\label{simulations}
In this section, we first compare the classical Black-Scholes model with the fractional stochastic volatility model  with different Hurst indexes. Taking European options as an example, we compare European option prices with different volatilities, maturities, and strike prices. Subsequently, we fix other parameters and adjust $\gamma$, apply Theorem \ref{thm:5} to approximate the European option price.

Firstly, we compare the prices of the four models corresponding to European options at different $\alpha$ and strike prices. In Table \ref{tab1} we illustrate the models. In Table \ref{tab2} and Fig. 1, we compare the option prices. Through the above numerical simulation we can clearly observe the volatility smile phenomenon.

\begin{table}[]
	{\footnotesize
		\caption{$T=1, X_{0}=50,\beta=0.5, \gamma=10.$}\label{tab1}
		\begin{center}
			\begin{tabular}{|c|c|c|} \hline
				Model & $dX_{t}$ & $dv_{t}$ \\ \hline		
				B-S & $d X_{t}=\alpha X_{t} d B_{t}$ &  \\
				OU & $d X_{t}=v_{t}X_{t} d B_{t}^{1}$ & $d v_{t}=\beta \left(\alpha-v_{t}\right) d t+\gamma d B_{t}^{2}$ \\ 
				FOU H=0.7 & $d X_{t}=v_{t}X_{t} d B_{t}$ & $d v_{t}=\beta \left(\alpha-v_{t}\right) d t+\gamma d B_{t}^{0.7}$ \\
				FOU H=0.9 & $d X_{t}=v_{t}X_{t} d B_{t}$ & $d v_{t}=\beta \left(\alpha-v_{t}\right) d t+\gamma d B_{t}^{0.9}$ \\ \hline
			\end{tabular}
		\end{center}
	}
\end{table}

\begin{table}[]
	{\footnotesize
		\caption{Option prrices with different $\alpha$ and $K$.}\label{tab2}
		\begin{center}
			\begin{tabular}{|ccccccccccc|}
				\hline
				K   & 5     & 10    & 15    & 20    & 25    & 30    & 35    & 40    & 45    & 50    \\ \hline
				\multicolumn{11}{|c|}{BS}                                                           \\ \hline
				$\alpha=$0.5 & 44.47 & 41.84 & 39.32 & 37.00 & 35.03 & 33.43 & 32.00 & 30.56 & 29.20 & 28.10 \\
				$\alpha=$1   & 44.37 & 43.75 & 43.25 & 42.69 & 42.16 & 41.81 & 41.51 & 41.16 & 40.80 & 40.54 \\
				$\alpha=$1.5 & 29.95 & 29.88 & 29.85 & 29.78 & 29.69 & 29.65 & 29.62 & 29.57 & 29.51 & 29.48 \\
				$\alpha=$2   & 5.31  & 5.29  & 5.24  & 5.21  & 5.20  & 5.20  & 5.20  & 5.19  & 5.19  & 5.18  \\
				$\alpha=$2.5 & 0.08  & 0.08  & 0.08  & 0.08  & 0.08  & 0.08  & 0.08  & 0.08  & 0.08  & 0.08  \\ \hline
				\multicolumn{11}{|c|}{OU}                                                           \\ \hline
				$\alpha=$0.5 & 30.99 & 29.92 & 28.96 & 28.09 & 27.34 & 26.69 & 26.11 & 25.59 & 25.09 & 24.65 \\
				$\alpha=$1   & 31.00 & 29.89 & 28.91 & 28.05 & 27.31 & 26.65 & 26.09 & 25.58 & 25.10 & 24.65 \\
				$\alpha=$1.5 & 30.73 & 29.62 & 28.62 & 27.76 & 27.02 & 26.39 & 25.86 & 25.36 & 24.89 & 24.46 \\
				$\alpha=$2   & 30.18 & 29.08 & 28.10 & 27.24 & 26.53 & 25.93 & 25.41 & 24.92 & 24.47 & 24.05 \\
				$\alpha=$2.5 & 29.30 & 28.21 & 27.27 & 26.45 & 25.78 & 25.19 & 24.70 & 24.25 & 23.82 & 23.41 \\ \hline
				\multicolumn{11}{|c|}{FOU H=0.7}                                                    \\ \hline
				$\alpha=$0.5 & 37.01 & 35.30 & 33.80 & 32.46 & 31.32 & 30.35 & 29.49 & 28.75 & 28.06 & 27.45 \\
				$\alpha=$1   & 38.08 & 36.08 & 34.49 & 33.14 & 32.00 & 31.01 & 30.21 & 29.52 & 28.85 & 28.25 \\
				$\alpha=$1.5 & 39.09 & 37.04 & 35.25 & 33.87 & 32.73 & 31.82 & 31.09 & 30.41 & 29.76 & 29.19 \\
				$\alpha=$2   & 39.85 & 37.91 & 36.14 & 34.71 & 33.63 & 32.78 & 32.08 & 31.42 & 30.80 & 30.24 \\
				$\alpha=$2.5 & 40.81 & 39.03 & 37.27 & 36.10 & 35.08 & 34.44 & 33.67 & 32.99 & 32.40 & 32.06 \\ \hline
				\multicolumn{11}{|c|}{FOU H=0.9}                                                    \\ \hline
				$\alpha=$0.5 & 31.46 & 29.50 & 27.72 & 26.09 & 24.60 & 23.25 & 22.03 & 20.97 & 20.03 & 19.21 \\
				$\alpha=$1   & 31.49 & 29.33 & 27.52 & 25.92 & 24.47 & 23.14 & 22.00 & 20.99 & 20.09 & 19.29 \\
				$\alpha=$1.5 & 31.72 & 29.57 & 27.64 & 26.06 & 24.65 & 23.43 & 22.36 & 21.40 & 20.54 & 19.78 \\
				$\alpha=$2   & 31.98 & 29.98 & 28.13 & 26.54 & 25.23 & 24.09 & 23.08 & 22.17 & 21.35 & 20.63 \\
				$\alpha=$2.5 & 32.79 & 31.03 & 29.17 & 27.93 & 26.73 & 25.89 & 24.81 & 23.88 & 23.12 & 22.67 \\ \hline
			\end{tabular}
		\end{center}
	}
\end{table}

\begin{figure}[]
	\centering
	\label{fig1}\includegraphics[width=1\textwidth]{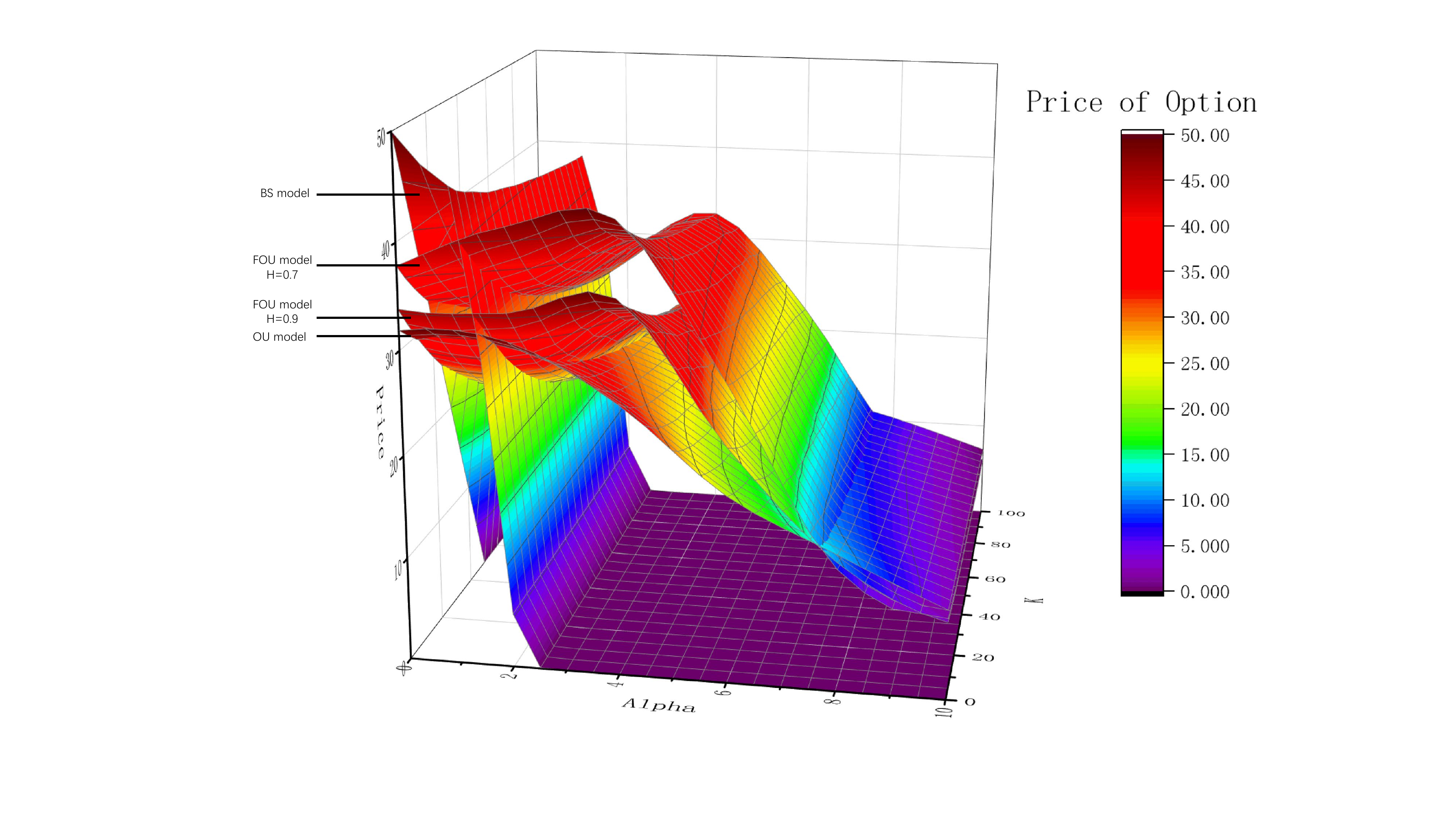}
	\caption{Comparison chart of option prrices with different $\alpha$ and $K$.}
\end{figure}

Secondly, we compare the prices of the three models corresponding to European options at different option maturities and strike prices. The forms of the models are the same as those in Table \ref{tab1} except for $\gamma =1$. In Table \ref{tab3} and Fig. 2, we compare the option prices. It can be seen that the price difference of options at different strike prices decreases as the option term increases.

\begin{table}[]
		\caption{Option prrices with different $T$ and $K$.}\label{tab3}
		\begin{center}
			\begin{tabular}{|ccccccccccc|}
				\hline
				K   & 5     & 10    & 15    & 20    & 25    & 30    & 35    & 40    & 45    & 50    \\ \hline
				\multicolumn{11}{|c|}{OU}                                                           \\ \hline
				T=1 & 45.96 & 41.71 & 37.89 & 34.49 & 31.47 & 28.80 & 26.44 & 24.37 & 22.55 & 20.95 \\
				T=4 & 43.90 & 41.89 & 40.26 & 38.88 & 37.68 & 36.63 & 35.69 & 34.85 & 34.08 & 33.38 \\
				T=8 & 41.93 & 41.09 & 40.44 & 39.89 & 39.42 & 38.99 & 38.61 & 38.27 & 37.95 & 37.65 \\ \hline
				\multicolumn{11}{|c|}{FOU H=0.7}                                                    \\ \hline
				T=1 & 45.15 & 40.84 & 36.98 & 33.52 & 30.45 & 27.74 & 25.35 & 23.25 & 21.41 & 19.81 \\
				T=4 & 42.06 & 39.99 & 38.29 & 36.85 & 35.59 & 34.49 & 33.52 & 32.64 & 31.86 & 31.14 \\
				T=8 & 30.04 & 29.13 & 28.42 & 27.83 & 27.31 & 26.86 & 26.45 & 26.08 & 25.73 & 25.42 \\ \hline
				\multicolumn{11}{|c|}{FOU H=0.9}                                                    \\ \hline
				T=1 & 45.14 & 40.80 & 36.90 & 33.42 & 30.33 & 27.60 & 25.19 & 23.08 & 21.24 & 19.62 \\
				T=4 & 40.82 & 38.67 & 36.87 & 35.33 & 33.99 & 32.81 & 31.77 & 30.83 & 30.00 & 29.24 \\
				T=8 & 27.15 & 26.09 & 25.24 & 24.53 & 23.91 & 23.36 & 22.87 & 22.43 & 22.03 & 21.66 \\ \hline
			\end{tabular}
		\end{center}
\end{table}

\begin{figure}[]
	\centering
	\label{fig2}\includegraphics[width=1\textwidth]{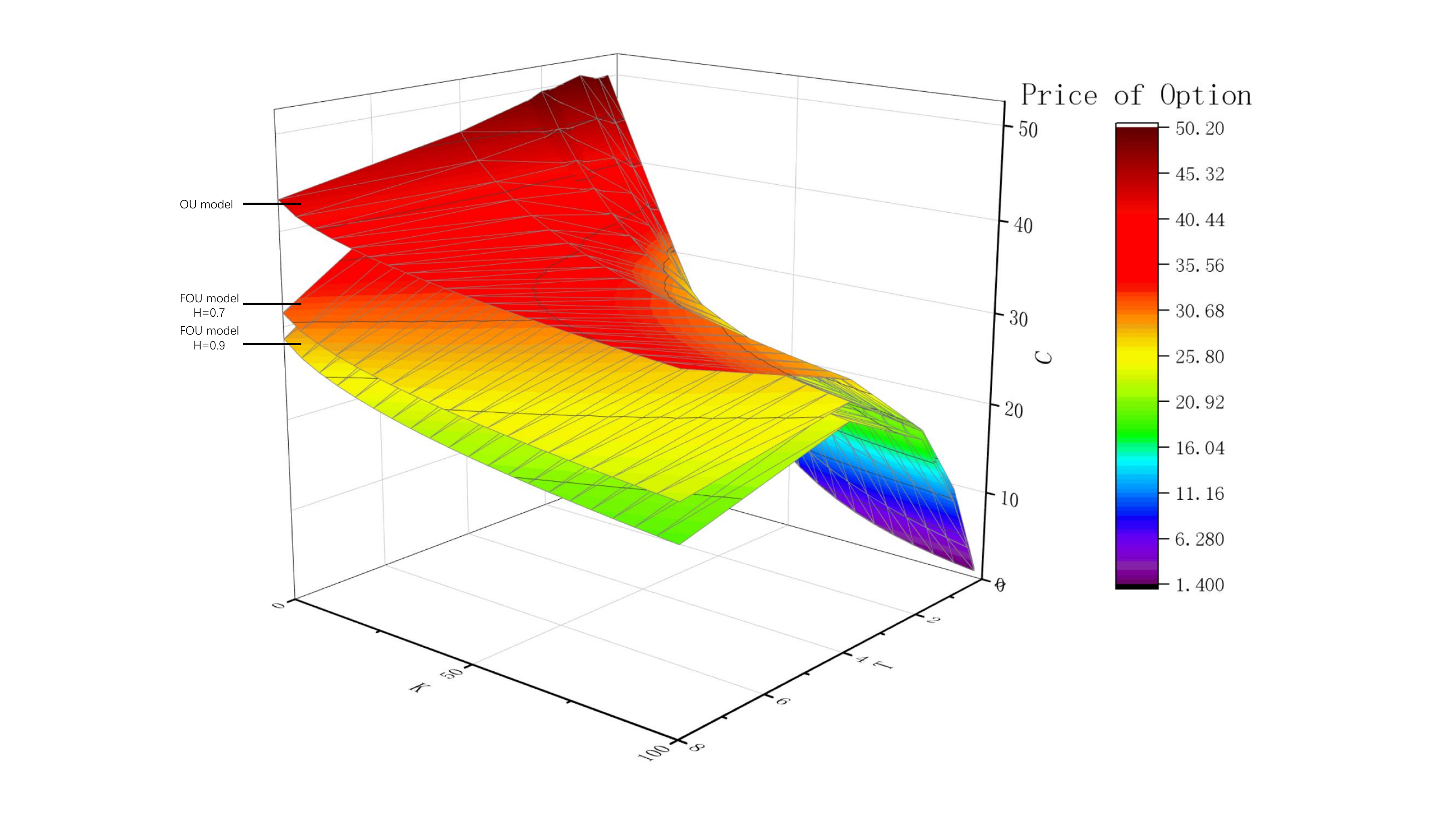}
	\caption{Comparison chart of option prrices with different $T$ and $K$.}
\end{figure}

Finally, we compare the prices of the fractional stochastic volatility model corresponding to European options under different $\gamma$. In Table \ref{tab4} we illustrate the models. In Table \ref{tab5} and Fig. 3, we compare option prices. When $\alpha$ takes on a specific range of values, the effect of long-range makes European options prices lower. The effect of long-range on derivatives prices decreases with the increase of $\alpha$.

\begin{table}[]
	{\footnotesize
		\caption{$T=1, X_{0}=50,\beta=0.5, H=0.9.$}\label{tab4}
		\begin{center}
			\begin{tabular}{|c|c|} \hline
				\multicolumn{2}{|c|}{$d X_{t}=v_{t}X_{t} d B_{t},d v_{t}=\beta \left(\alpha-v_{t}\right) d t+\gamma d B_{t}^{H}.$} \\ \hline
				Model & $\gamma$ \\ \hline
				FOU-I & 10   \\
				FOU-II & 0.1  \\ 
				FOU-III & 0.001 \\ \hline
			\end{tabular}
		\end{center}
	}
\end{table}

\begin{table}[]
		\caption{Comparison chart of option prrices with different $\alpha$ and $K$.}\label{tab5}
		\begin{center}
			\begin{tabular}{|ccccccccccc|}
				\hline
				K   & 5     & 10    & 15    & 20    & 25    & 30    & 35    & 40    & 45    & 50    \\ \hline
				\multicolumn{11}{|c|}{FOU-I}                                                        \\ \hline
				$\alpha=$0.5 & 31.46 & 29.50 & 27.72 & 26.09 & 24.60 & 23.25 & 22.03 & 20.97 & 20.03 & 19.21 \\
				$\alpha=$1   & 31.49 & 29.33 & 27.52 & 25.92 & 24.47 & 23.14 & 22.00 & 20.99 & 20.09 & 19.29 \\
				$\alpha=$1.5 & 31.72 & 29.57 & 27.64 & 26.06 & 24.65 & 23.43 & 22.36 & 21.40 & 20.54 & 19.78 \\
				$\alpha=$2   & 31.98 & 29.98 & 28.13 & 26.54 & 25.23 & 24.09 & 23.08 & 22.17 & 21.35 & 20.63 \\
				$\alpha=$2.5 & 32.79 & 31.03 & 29.17 & 27.93 & 26.73 & 25.89 & 24.81 & 23.88 & 23.12 & 22.67 \\ \hline
				\multicolumn{11}{|c|}{FOU-II}                                                       \\ \hline
				$\alpha=$0.5 & 43.35 & 39.07 & 34.91 & 30.73 & 26.25 & 21.52 & 17.31 & 14.14 & 11.29 & 8.91  \\
				$\alpha=$1   & 43.41 & 39.07 & 35.22 & 31.50 & 27.44 & 23.51 & 20.62 & 17.98 & 15.54 & 13.57 \\
				$\alpha=$1.5 & 43.63 & 39.57 & 35.78 & 32.46 & 29.03 & 26.35 & 24.04 & 21.81 & 19.80 & 18.13 \\
				$\alpha=$2   & 43.83 & 40.22 & 36.82 & 33.79 & 31.25 & 29.11 & 27.22 & 25.41 & 23.76 & 22.38 \\
				$\alpha=$2.5 & 44.67 & 41.62 & 38.48 & 36.19 & 34.12 & 32.69 & 31.07 & 29.53 & 28.16 & 27.30 \\ \hline
				\multicolumn{11}{|c|}{FOU-III}                                                      \\ \hline
				$\alpha=$0.5 & 42.92 & 38.62 & 34.44 & 30.24 & 25.77 & 21.03 & 16.81 & 13.63 & 10.78 & 8.40  \\
				$\alpha=$1   & 42.96 & 38.50 & 34.61 & 30.88 & 26.82 & 22.89 & 19.99 & 17.34 & 14.90 & 12.93 \\
				$\alpha=$1.5 & 43.42 & 39.23 & 35.34 & 32.02 & 28.59 & 25.91 & 23.60 & 21.36 & 19.36 & 17.69 \\
				$\alpha=$2   & 44.00 & 40.35 & 36.88 & 33.82 & 31.30 & 29.17 & 27.28 & 25.48 & 23.83 & 22.47 \\
				$\alpha=$2.5 & 45.64 & 42.66 & 39.44 & 37.29 & 35.23 & 33.93 & 32.26 & 30.69 & 29.33 & 28.59 \\ \hline
			\end{tabular}
		\end{center}
\end{table}

\begin{figure}[]
	\centering
	\label{fig3}\includegraphics[width=1\textwidth]{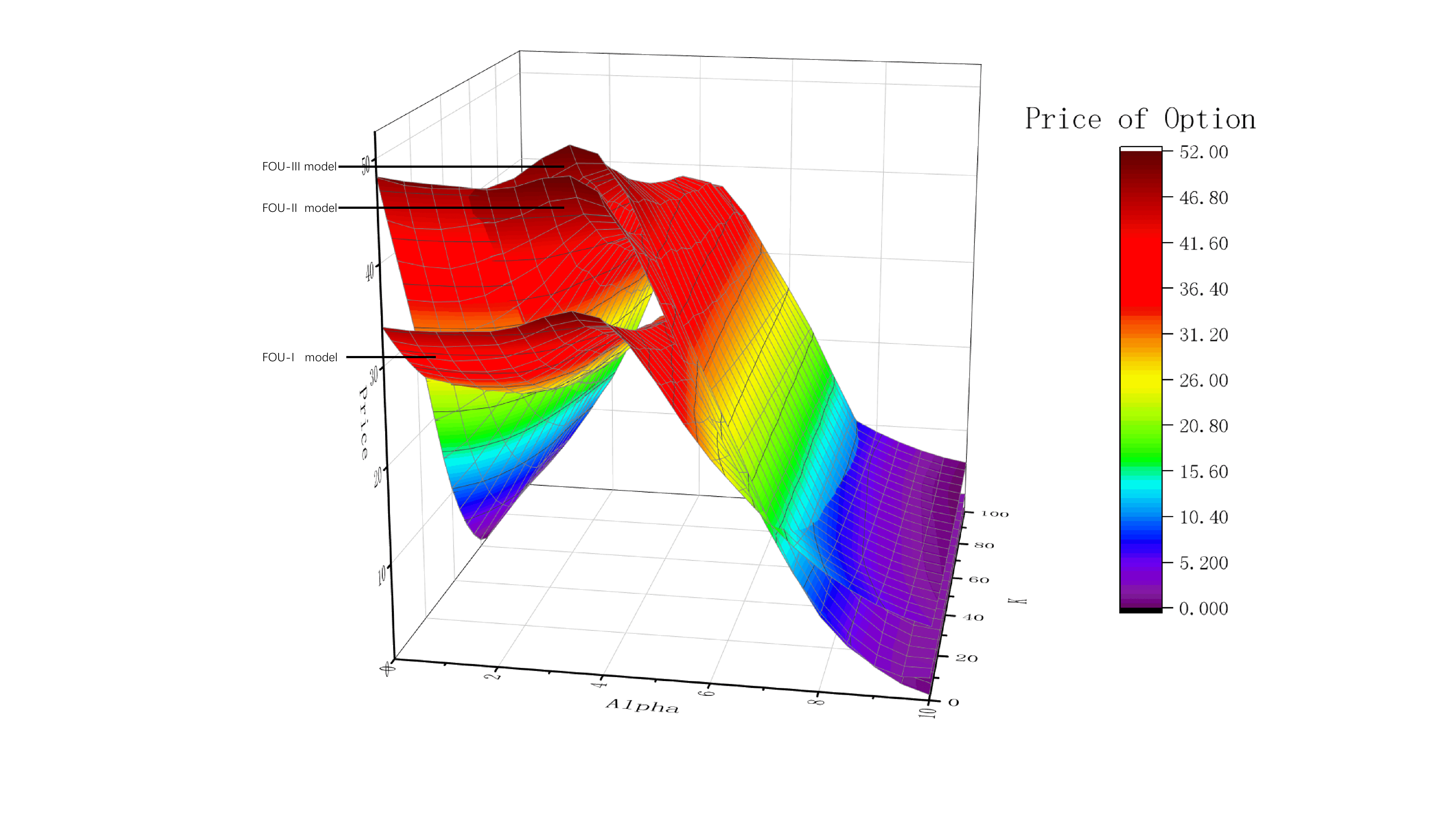}
	\caption{Comparison chart of option prrices with different $\alpha$ and $K$.}
\end{figure}

\section{Conclusions}
\label{sec:conclusions}
In this paper, we investigate the problem of pricing derivatives under a fractional stochastic volatility model. We obtain a method for approximating the pricing of derivatives where the stochastic volatility can be composed of deterministic functions of time and the fractional Ornstein-Uhlenbeck process. Some fractional stochastic volatility models can be generalized to this type of problem. As an example, we give an approximate pricing expression and numerical simulation of a European option under the fractional Stein-Stein model. The numerical simulation results show a clear volatility smile phenomenon, and the price difference between European options at different strike prices shrinks as the strike period increases. Numerical simulation results also demonstrate the effect of long-range on derivative prices.

\begin{acknowledgements}
This work is partially supported by the National Science Foundation of China (grant no. 11871244).
\end{acknowledgements}

%
%



\end{document}